# Tuning the thermal conductivity of Si membrane using nanopillars: from crystalline to amorphous pillars

Lina Yang[1*], Yixin Xu[2], Xianheng Wang[1], Yanguang Zhou[2,3*]

[1] School of Aerospace Engineering, Beijing Institute of Technology, Beijing 100081, China

[2] Department of Mechanical and Aerospace Engineering, The Hong Kong University of Science and Technology, Clear Water Bay, Kowloon, Hong Kong Special Administrative Region， China

[3] HKUST Shenzhen-Hong Kong Collaborative Innovation Research Institute, Futian, Shenzhen, Guangdong, China

Email: yangln@bit.edu.cn (L. Yang); maeygzhou@ust.hk (Y. Zhou)

## Abstract

Tuning thermal transport in nanostructures is essential for many applications, such as thermal management and thermoelectrics. Nanophononic metamaterials (NPM) have shown great potential for reducing thermal conductivity. In this work, the thermal conductivity of NPM with crystalline Si (c-Si) pillar, crystalline Ge (c-Ge) pillar and amorphous Si (a-Si) pillar are systematically investigated by molecular dynamics method. The analyses of phonon dispersion and spectral energy density show that phonon dispersions of Si membrane are flattened due to local resonant hybridization induced by both crystalline and amorphous pillar. In addition, a-Si pillar can cause larger reduction of thermal conductivity compared with c-Si pillar. Specifically, when increasing the atomic mass of atoms in pillars, the thermal conductivity of NPMs with crystalline pillar is increased because of the weakened phonon hybridization, however, the thermal conductivity of NPMs with amorphous pillar is almost unchanged. The analyses of the reduction of thermal conductivity shows that both resonant hybridization and scattering mechanism are important in NPMs with crystalline pillar, while scattering mechanism dominates in NPMs with amorphous pillar and NPMs with short crystalline pillar. The results of this work can provide meaningful insights on controlling thermal transport in NPMs by choosing the materials and atomic mass of pillars for specific applications.

## 1. Introduction

Controlling thermal transport in nanostructures has attracted significant attention in recent years due to its promising applications such as thermal management[1,2] and thermoelectric energy conversion[3-5]. For instance, in the pursuit of high-efficiency thermoelectrics, a promising avenue of the strategies lies in the development of nanophononic metamaterials (NPM) with low thermal conductivity ($\kappa$).[6-8] NPMs enable phonon resonant hybridizations between the vibrational modes of nanoresonators and the phonon modes of the host medium, leading to enhanced control over thermal conductivity.[9-12] Meanwhile, previous studies found the pillared silicon (Si) membrane can reduce the $\kappa$ by two orders of magnitudes[9,13]. Unlike nanoscale phononic crystals[14,15], NPM does not require periodicity in the arrangement of resonators to achieve resonant hybridizations, which renders NPM highly robust to disorder in the arrangement of resonators[9]. Additionally, as the resonators of NPM are located outside the membrane[7,13,16], electron transport in these materials remains almost unaffected, which is highly desirable for developing high-efficiency thermoelectrics.

Substantial efforts have been aimed at comprehending the thermal transport properties of NPMs and the underlying mechanisms since the NPM was investigated by Davis and Hussein in 2014 [6]. For instance, Anufriev et al. found that the $\kappa$ of nanobeams with aluminum (Al) nanopillar is about 20% smaller compared with the pristine nanobeams, which is mainly caused by phonon scatterings at the pillar/beam interface due to the intermixing of Al and Si atoms[17]. Maire et al. showed that the cross-sections control the $\kappa$ of nanowires with fishbone nanostructures, and the periodic wings further reduce the $\kappa$[18]. Later analyses found that the reduction of κ is mainly caused by the periodicity of wings which can flatten the phonon dispersions, rather than by local resonances[19]. By conducting atomistic modeling and experiments, Neogi et al. found that the reduction of $\kappa$ in ultrathin suspended silicon membranes is mainly controlled by surface scatterings because the rough layers of native oxide at surfaces limit the mean free path of thermal phonons below 100 nm.[20] Similarly, Huang et al. reported that the $\kappa$ of suspended silicon membranes with nanopillars is controlled by incoherent phonon scatterings causing less than 16% reduction of $\kappa$, which is examined by comparing the results of Monte Carlo simulation and experimental measurements.[21] Recently, the device-scale suspended silicon membranes with GaN nanopillars were fabricated, and experimental measurements found that the nanopillars cause up to 21% reduction of $\kappa$, meanwhile, the power factor remains unaffected[16]. The effect of rough surface (sawtooth-shaped surface and square-shaped surface), amorphous surface, and smooth surface on phonon reflection were studied by atomistic wave-packet simulations[22]. It is found that the reflected wave-packet distribution does not follow the well-known Ziman model but shows a nonmonotonic dependence on the depth of the surface roughness, while the amorphous surface can absorb the incident wave packet, and then reflect it diffusively.

Besides these experimental works, different types of NPMs have been studied theoretically. Xiong et al. found that combining a designed resonant structure with alloying can lead to extremely low $\kappa$ in Si nanowires, because the local resonances greatly reduce the phonon group velocities and mean free paths in the low frequency (<4 THz) range, concurrently, alloy scatterings impede high frequency phonons.[7] Later, Zhang et al. designed a Si nanowire with helical wall which is more effective on reducing $\kappa$ compared with straight nanowalls and nanopillars. They demonstrated the resonant hybridization and mode localization in the helical walls by analyzing the phonon dispersions and phonon spatial distributions, respectively.[23] Further, it was found that introducing imperfections such as vacancy defects, mass mismatch, and alloy disorder in the pillars can weaken the local resonant hybridization, leading to a higher $\kappa$ compared with that of the pristine NPMs[24,25]. For instance, the $\kappa$ of pillared graphene nanoribbon (GNR) is increased from ~47 W/mK to ~63 W/mK by increasing the atomic mass of atoms in pillars.[24] Although previous works have reported that introducing nanopillars will affect $\kappa$ through several mechanisms[11] such as local resonance[13], phonon interference due to pillar periodicity[19] and diffuse phonon scatterings[17,20,21], tuning $\kappa$ of NPMs by designing different kinds of resonators is less investigated, and the corresponding phonon transport behaviors remain unclear.

In this work, the effect of the crystalline Si (c-Si) pillar, amorphous Si (a-Si) pillar and crystalline Ge (c-Ge) pillar on regulating $\kappa$ of Si membrane is investigated by equilibrium molecular dynamics simulations (EMD). In addition, the mass of atoms in pillars is tuned to manipulate the $\kappa$ of NPMs with c-Si and a-Si pillars. The phonon local resonant hybridizations in NPMs are systematically analyzed by calculating phonon dispersions and spectral energy distribution (SED). Further, the spectral thermal conductivity is also quantified. The results of our work are expected to provide insights into the interplay between the resonators of NPMs and phonon transport.

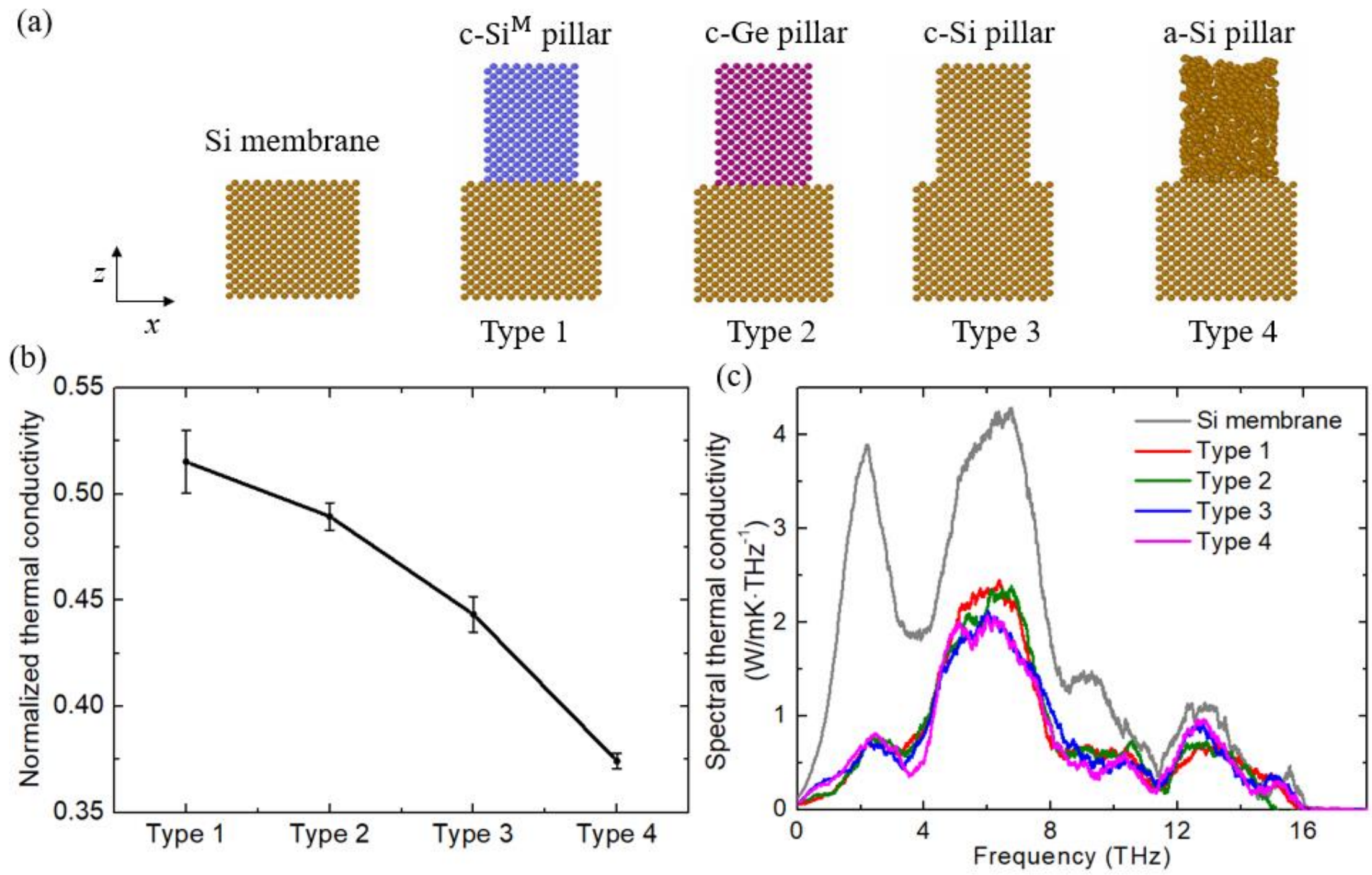

Figure 1. (a) The structure Si membrane, NPM with c-Si$^M$ pillar (Type 1), NPM with c-Ge pillar (Type 2), NPM with c-Si pillar (Type 3) and NPM with a-Si pillar (Type 4). (b) Normalized thermal conductivity of NPM from Type 1 to Type 4. The $\kappa$ of NPM is normalized by the $\kappa$ of Si membrane. (c) Spectral thermal conductivity of Si membrane and NPMs versus frequency. In (b) and (c), the value of M is set as MGe for Type 1.

## 2. Model and Method

To study the $\kappa$ in NPMs, Si membrane with c-Si pillar, c-Ge pillar, and a-Si pillar in Figure 1 (a) are created, which are noted as Type 3, Type 2, and Type 4, respectively. When the atomic mass of atoms in the c-Si pillar is changed to M in amu unit, it is noted as c-Si$^M$ pillar. The NPM with c-Si$^M$ pillar is represented by Type 1. Here, the atomic mass of Si and Ge atoms is denoted by MSi and MGe, respectively. If M equals MSi, the c-Si$^M$ pillar represents the c-Si pillar. In Figure 1 (b) and (c), the value of M is set as MGe for Type 1.

$La = 5.431$Å and $L_{Ge} = 5.66$ Å are the lattice constant of bulk Si and Ge, respectively. The unit cell of Si membrane is $6La \times 6La \times 6La$. The size of the c-Ge pillar is $4L_{Ge} \times 4L_{Ge} \times 6L_{Ge}$, and the size of other pillars is $4La \times 4La \times 6La$. The images of the system structures are created with OVITO[26], version 2.9.0, in this paper.

The thermal conductivity of NPM is calculated by EMD method. In the simulations, the periodic boundary condition is applied in the *x* and *y* directions, and the free boundary condition is applied in the *z* direction. Tersoff potential is used to describe the interaction between Si atoms[27] and Ge atoms. The couplings between Si and Ge atoms are obtained by following the combination rule [27], which has been applied to describe the Si and the SiC system. All molecular dynamics simulations are performed using LAMMPS (large-scale atomic/molecular massively parallel simulator)[28], and the temperature is set as 300 K. Thermal conductivity is calculated from the Green-Kubo formula[29],

$$\kappa = \frac{1}{2Vk_BT^2}\int_0^\infty \langle J(0) \cdot J(\tau)\rangle \, d\tau \tag{1}$$

where $k_B$ is the Boltzmann constant, *V* is the system volume, *T* is the temperature and *J* is the heat current. The time step in EMD simulations is set as 0.5 fs. Firstly, the NPM system is equilibrated at 300 K using canonical ensemble with Langevin heat reservoir for $6\times10^5$ steps (300 ps). Then, the heat current is recorded at each step during the EMD simulations under microcanonical ensemble (NVE) for $8\times10^6$ steps (4 ns). In this work, the time averaging supplemented by twenty simulation cases with different initial conditions is used to calculate the thermal conductivity, which has been used in many research studies[29-31], and the statistical errors are obtained by calculating the standard deviation.

The a-Si pillar is obtained by the melt-quenching method[32-35]. Initially, the Si membrane with the c-Si pillar is created. The atoms in the Si membrane are fixed, and the atoms in the c-Si pillar are melted at 3600 K by Nosé-Hoover thermostat for 0.5 ns, then the pillar is quickly quenched to 1000 K at the rate of 860 K/ps. Finally, the atoms in the pillar are annealed at 1000 K for 0.5 ns, then quickly quenched to 20 K at the rate of 160 K/ps. To determine the atomic configuration of the a-Si pillar, an equilibration simulation is first performed for the a-Si pillar for 2.5 ns in an NVT ensemble, and then the steady atomic positions are calculated by averaging the time-dependent atomic positions during the next 2.5 ns.

To characterize the phonon transport in NPMs, the phonon dispersions of NPMs and Si membranes are calculated by lattice dynamics (LD) implemented in GULP[36]. The SED[37-39] is obtained directly from the MD simulations by recording the velocities of atoms. The SED expression $\Phi'$ is given by

$$\Phi'(k,\omega) = \frac{1}{4\pi\tau_0}\sum_{\alpha}^{3}\sum_{b=1}^{n}\frac{m_b}{N}\left|\sum_{1}^{N}\int_{0}^{\tau_0}\dot{u}_\alpha(l,b,t)exp\Theta dt\right|^2 \quad (2)$$

where $\dot{u}_\alpha$ is the $\alpha$th component of the velocity of the $b$th atom in the $l$th unit cell at time *t*, and $\Theta = i[\vec{k}\cdot\vec{r_0}(l,b) - \omega t]$. $m_b$ is the mass of the $b$th atom. $\tau_0$ is the simulation time, $\vec{r_0}$ is the equilibrium position vector of the $l$th unit cell, and $\omega$ is the angular frequency. Here, the unit of $\Phi'$ is J·ps. *N* is the total number of unit cells, and *n* is the number of atoms in the unit cell. In the SED analyses, the supercell of NPM consists of 40 unit cells in the x direction and 1 unit cell in the y and z directions. Here, the MD simulations are performed at 300 K for $2\times10^6$ time steps to extract the atomic velocities.

To quantify the spectral phonon transmission function, we further calculate the spectral thermal conductivity using nonequilibrium molecular dynamics simulations (NEMD), in which the spectral heat current can be calculated by [40,41]

$$Q(\omega) = 2\sum_{i\in left}\sum_{j\in right} Re\left[\int_{-\infty}^{+\infty}\left\langle\frac{\partial U_j}{\partial\vec{r}_i}\bigg|_\tau \vec{v}_i(0) - \frac{\partial U_i}{\partial\vec{r}_j}\bigg|_\tau \vec{v}_j(0)\right\rangle e^{-i\omega\tau}d\tau\right] \quad (3)$$

where $U_j$ is the potential energy of atom *j*, $\vec{v}_i$ is the velocity of atom *i* and $\langle\rangle$ denotes the time average in MD simulations. The atomic velocities $\vec{v}_i$ and the atomic potential partial function $\frac{\partial U_j}{\partial\vec{r}_i}$ are updated every 20 steps, i.e., 10 fs. The size effects and the configuration of NEMD simulations are shown in Figure S5 in the Supplementary Material[42]. After that, by assuming the same temperature gradient $\nabla T$ over all the phonons, the spectral thermal conductivity is then calculated by Fourier's law, i.e., $\kappa(\omega) = Q(\omega)/\mathrm{A}\nabla T$ in which A denotes the system cross-section area and $\nabla T$ is the temperature gradient along the in-plane direction.

## 3. Results

### 3.1 Thermal conductivity of NPMs

The thermal conductivity of NPMs in Figure 1 (a) is calculated by EMD method through Eq.(1). In Figure 2 (b) and (c), the value of M is set as MGe for c-Si$^{M}$ pillar in Type 1. For comparison, the thermal conductivity of bulk Si and Si membranes in Figure 1 (a) is also computed. The simulation cell of bulk Si is set as $6La\times6La\times6La$ with periodic boundary conditions in all three spatial directions to overcome the size effect[29,43,44]. The calculated $\kappa$ is 238.4±3.6 W/m-K and 40.6±0.6 W/m-K for bulk Si and Si membrane at 300 K, respectively, which is consistent with the prediction in Ref.[45] and Ref.[20]. The thermal conductivity as a function of time is shown in Figure S1 in the Supplementary Material[42]. The thermal conductivity of the Si membrane calculated by the NEMD method is 28.19 W/mK at T = 300 K in Figure S5 in the Supplementary Material[42], which is also higher than experimental results (~10 W/m-K for a membrane with a thickness of ~10 nm[46]). It is noted that there exists a gap between the MD results and the experimental measurements[47], which can be caused by the intrinsic phonon distribution and the accuracy of empirical potentials. On one hand, all the vibrational modes are assumed to be fully occupied in the MD simulations, and they follow the classical limit of Boltzmann distributions[48], as a result, the MD simulations overestimate the heat capacity and the corresponding thermal conductivities. On the other hand, the accuracy of MD results strongly relies on the empirical potential. Here, the popular potential for the Si (Ge) systems [27] is applied, which has been widely used to study thermal transport in Si-based materials, including bulk Si[29], Si/Ge superlattices[49], and Si membranes[50]. However, this discrepancy will not severely affect the comparison of NPMs and Si membranes, since the same simulation methods are used and the fundamental mechanisms should be independent of the potential. The $\kappa$ of NPMs is normalized by the $\kappa$ of Si membrane, which is shown in Figure 1 (b). The normalized $\kappa$ of NPM with c-Si$^{M}$ pillar and c-Ge pillar is larger than that of NPM with c-Si pillar, while a-Si pillar has a stronger effect on the reduction of $\kappa$ compared c-Si pillar.

To further understand the reduction $\kappa$ of NPMs in Figure 1 (b), the spectral $\kappa$ (Figure 1 (c)) is investigated. Compared with the Si membrane, the spectral $\kappa$ of NPMs is reduced in a wide range of frequencies from 0 to 14 THz. At the low-frequency range (< 2.0 THz), the c-$Si^M$ pillar and the c-Ge pillar cause slightly smaller spectral $\kappa$ compared with the c-Si pillar. However, from 5 to 7.5 THz, NPMs with c-$Si^M$ pillar and c-Ge pillar have larger spectral $\kappa$, which leads to the larger $\kappa$ compared with NPM with c-Si pillar. Furthermore, the contribution of low-frequency phonons (< 2.0 THz) to spectral $\kappa$ is almost equivalent in the a-Si pillared NPM and the c-Si pillared NPM. However, the spectral $\kappa$ of NPM with a-Si pillar is smaller for frequency ranging from 3.5 to 4.5 THz and 7.5 to 10 THz, which leads to the smaller $\kappa$ compared with NPM with the c-Si pillar.

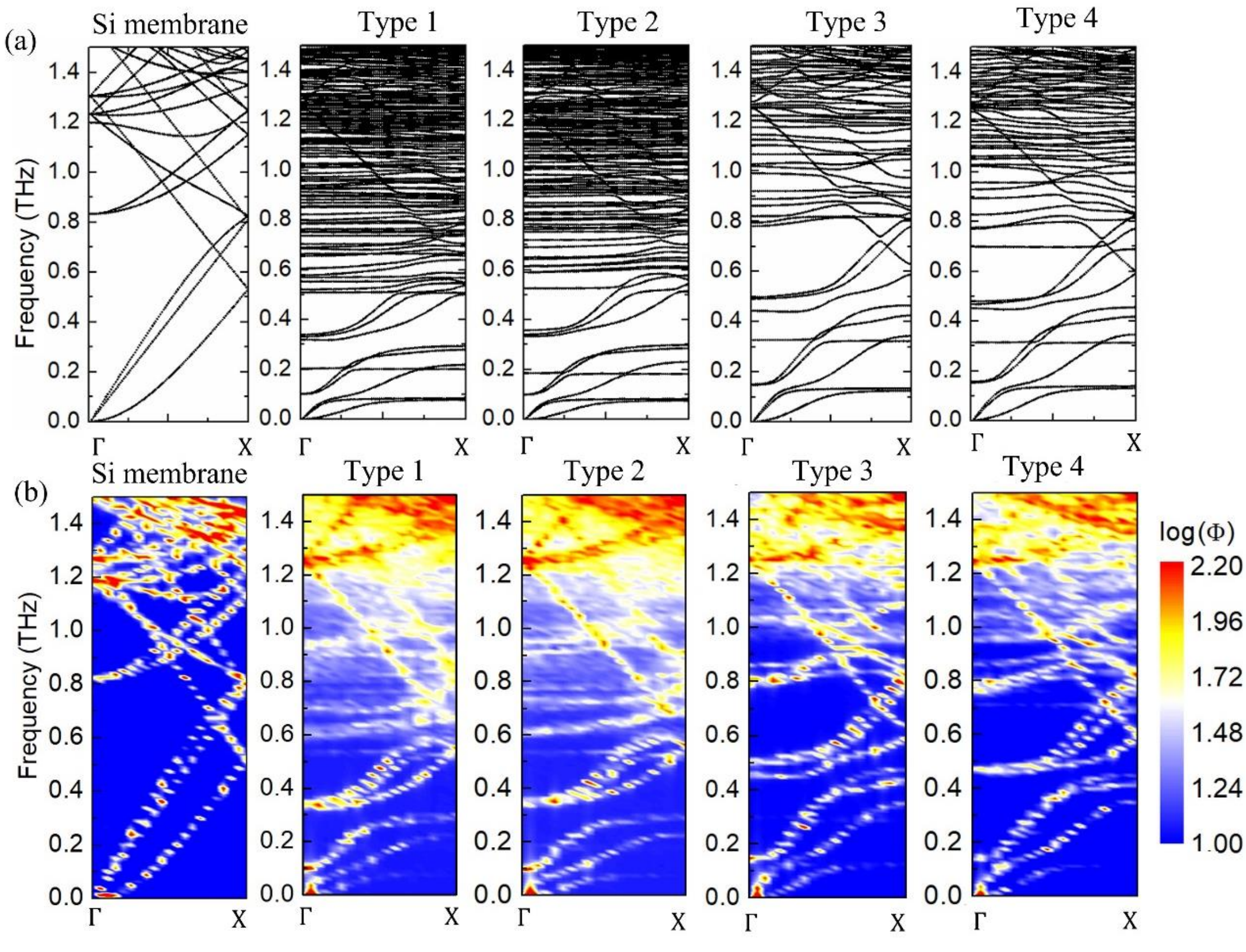

Figure 2 (a) Phonon dispersions of Si membrane and NPMs calculated by lattice dynamics. (b) SED spectrum of Si membrane and NPMs only considering atoms in the Si membrane. The meaning of Type 1 to Type 4 is the same as in Figure 1.

It was reported that the phonon hybridization between the host medium and the resonant surface structures played an essential role in reducing the thermal conductivity of nanostructures[7,19,25]. Here, the phonon hybridizations in the NPMs are analyzed by calculating their phonon dispersion using the lattice dynamics method implemented in GULP[36], shown in Figure 2 (a). In addition, the SED spectrum of the Si membrane and the Si membrane in NPMs, as shown in Figure 2 (b), are also calculated. The SED spectrum of NPMs considering all atoms in the system is shown in Figure S2 in the Supplementary Material[42]. Because phonon modes with high frequency are difficult to distinguish in the SED spectrum, we only consider the low-frequency phonon modes ($\leq$ 1.5 THz) here. The results show that introducing the nanopillars can affect the low-frequency phonon modes of Si membrane, for example, the first resonant hybridization causing the flattened bands happens at ~0.1 THz for Type 1 and Type 2 NPMs and at ~0.15 THz for Type 3 and Type 4 NPMs, which leads to a reduction of $\kappa$ of NPM compared with the pure Si membrane. In addition, the phonon dispersions of Type 1 and Type 3 are close to that of Type 2 and Type 4, respectively. However, the lattice mismatch at the interface between the c-Ge pillar and Si membrane causes stronger phonon scatterings, therefore, NPM with the c-Ge pillar can have a smaller $\kappa$ than NPM with the c-$Si^M$ pillar. Similarly, the roughness of the amorphous pillar also causes stronger phonon scatterings at the interface between the a-Si pillar and Si membrane, therefore, NPM with the a-Si pillar can have a smaller $\kappa$ than NPM with the c-Si pillar.

### 3.2 Anomalous thermal conductivity of NPMs

The frequency of vibrational modes of the pillar can be controlled by varying its atomic mass. Additionally, it was reported that changing the atomic mass of the pillar can also weaken the resonant hybridization and abnormally increase the $\kappa$[25]. To further study the effect of the atomic mass, the thermal conductivity of NPM with c-$Si^M$ pillar and a-$Si^M$ pillar is calculated, where the value of M is varied from 0.5 MSi to 10 MSi. Here, when the atomic mass of atoms in pillars is changed to M in amu unit, it is denoted as c-$Si^M$pillar and a-$Si^M$ pillar. The thermal conductivity of NPM is normalized by $\kappa$ of the Si membrane, which is shown in Figure 3 (a), the $\kappa$ of NPMs versus time are shown in Figure S3 in the Supplementary Material[42]. For NPM with the c-$Si^M$ pillar, $\kappa$ is the smallest when M equals MSi, and largely increased as the M increases from MSi to 10 MSi, which is consistent with the findings of graphene NPM with larger mass pillar.[24,25] Similarly, the NPM with a-$Si^M$ pillar has a larger $\kappa$ when M is decreased from MSi to 0.5 MSi. However, its $\kappa$ is almost unchanged as M increases from MSi to 10 MSi. The different trends of $\kappa$ of NPMs as M increases in Figure 3 (a) indicates that the mechanism of reduction of $\kappa$ is different for NPMs with c-$Si^M$ pillar and NPMs with a-$Si^M$ pillar.

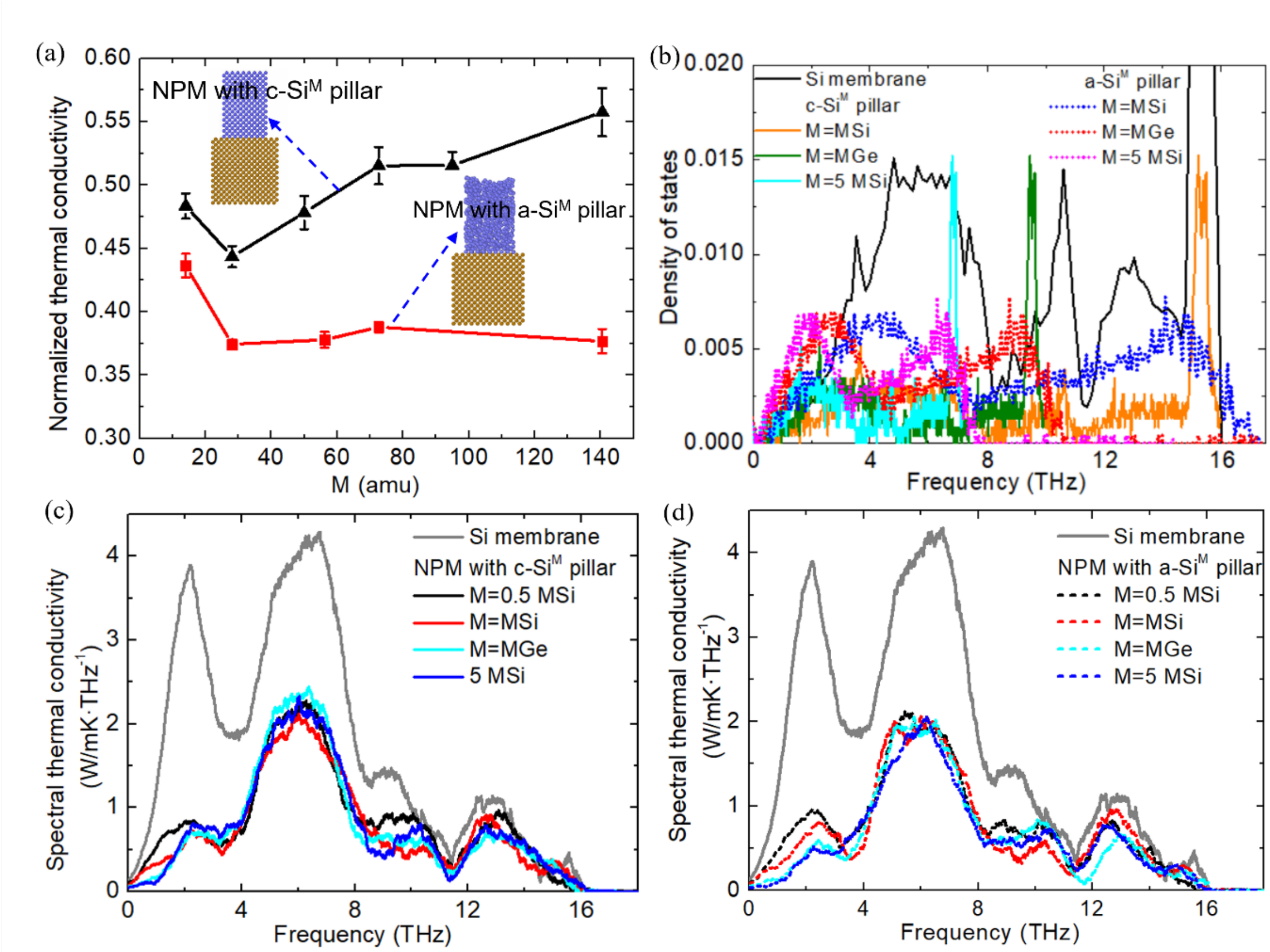

Figure 3 (a) Normalized thermal conductivity of NPM with c-$Si^M$ pillar and a-$Si^M$ pillar versus the value of M. Here, when the atomic mass of atoms in pillars is changed to M in amu, it is denoted as c-$Si^M$ pillar and a-$Si^M$ pillar. The thermal conductivity of NPM is normalized by $\kappa$ of the Si membrane. (b) Density of states (DOS) versus frequency for c-$Si^M$ pillar and a-$Si^M$ pillar. (c) Spectral thermal conductivity of NPM with c-$Si^M$ pillar. (d) Spectral thermal conductivity of NPM with a-$Si^M$ pillar.

To understand the different behavior of $\kappa$ between NPM with c-$Si^M$ pillar and a-$Si^M$ pillar as M increases from MSi to 10 MSi, the density of states (DOS) in Figure 3 (b) and spectral $\kappa$ in Figure 3 (c) and (d) are calculated. Figure 3 (c) shows that for NPM with the c-$Si^M$ pillar, the spectral $\kappa$ contributed by the low-frequency phonon (<2 THz) is decreased due to the stronger hybridization at the low-frequency as M increases (shown in Figure 4), while, the spectral $\kappa$ contributed by phonon with frequency from 4.5 to 7.5 THz is increased, leading to the increase of the $\kappa$. Figure 3 (d) shows that in NPM with a-$Si^M$ pillar, the spectral κ decreases as M increases for low frequency phonons (< 3.5 THz), and for other frequency ranges, it does not show obvious trends as M increases.

The DOS in Figure 3 (b) shows that the vibrational frequencies of the c-$Si^M$ pillar and a-$Si^M$ pillar are suppressed toward low frequency as M increases. For example, the range of frequency of the vibrational modes is suppressed to ∼7.0 THz and 5.0 THz as M increases from MSi to 5 MSi and 10 MSi, respectively. Therefore, there will be fewer resonant hybridizations above ∼7.0 THz and ∼5.0 THz in the NPMs, which means that the range of frequency of resonant hybridization is reduced as M increases. In addition, changing the atomic mass of the pillar can weaken the resonant hybridizations[25]. Therefore, the $\kappa$ of NPM with c-$Si^M$ pillar is increased as M increases due to the weakened resonant hybridizations and reduced frequency range of resonant hybridizations. The accumulated thermal conductivity of Si membrane versus frequency is shown in Figure S6 (a), and the DOS of c-$Si^{10MSi}$ pillar and a-$Si^{10MSi}$ pillar is shown in Figure S6 (b) in the Supplementary Material[42]. The contribution of phonons with frequency smaller than 5.0 THz is 38% in Si membrane in Figure S6 (a). If these phonons were completely impeded because of the resonant hybridization mechanism in NPM with c-$Si^{10MSi}$ pillar, the reduction of $\kappa$ would be 38%, additionally, these phonons could not be completely impeded which is shown in Figure 4, therefore, the reduction of $\kappa$ of NPM with c-$Si^{10MSi}$ pillar due to resonant hybridization should be smaller than 38%. However, the $\kappa$ of NPM with c-$Si^{10MSi}$ pillar is reduced 39%, which implies that both resonant hybridization mechanism and scattering mechanism have substantial contribution to the reduction of $\kappa$ of NPM with c-Si pillar. Similarly, a previous work quantified the particle and wave effects on the reduction of $\kappa$ of one type of NPMs (the Si nanowire cross junction) by combining Monte Carlo and atomic Green's function methods, it was found that particle effect is quite significant as well and can contribute as much as 39% to the total reduction of $\kappa$, highlighting the importance of mutual particle and wave effect in the NPMs[51].

On the other hand, the $\kappa$ of NPM with a-Si$^M$ pillar is almost unchanged as M increases, which implies that the phonon scatterings at the interface between the a-Si$^M$ pillar and the Si membrane dominate the suppression of phonon transport.

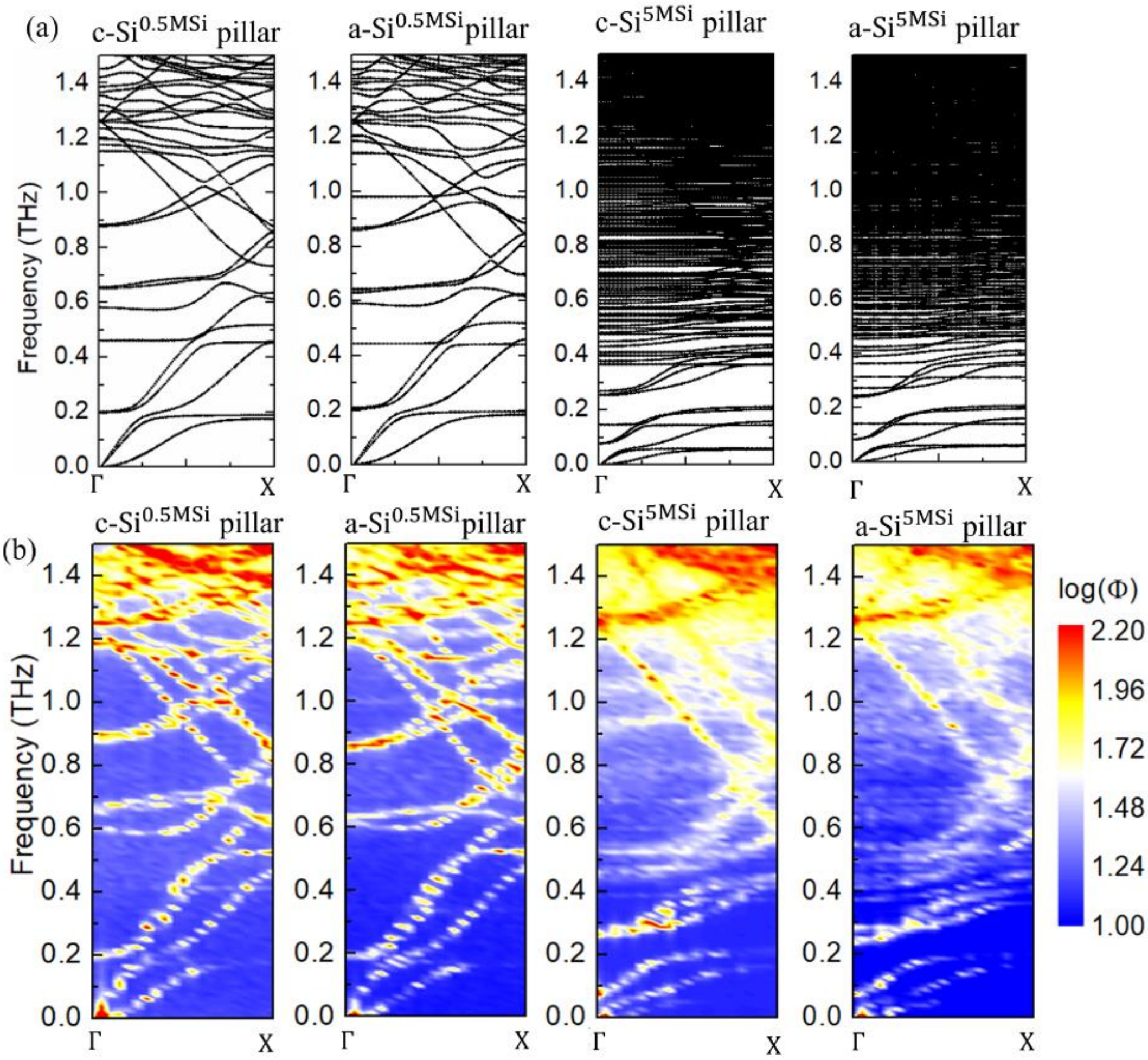

Figure 4 (a) Phonon dispersions of NPMs with c-Si$^M$ pillar and a-Si$^M$ pillar calculated by lattice dynamics. (b) SED spectrum of NPMs with c-Si$^M$ pillar and a-Si$^M$ pillars, in which only the atoms in Si membrane are considered. The value of M is set as 0.5 MSi and 5 MSi to compare the change of phonon dispersions. Here, c-Si$^{0.5MSi}$ and a-Si$^{0.5\ MSi}$ pillar represent c-Si$^M$ and a-Si$^M$ pillar with M=0.5 MSi, and c-Si$^{5MSi}$ and a-Si$^{5MSi}$ pillar represents c-Si$^M$ and a-Si$^M$ pillar with M=5 MSi.

Furthermore, the value of M is set as 0.5 MSi and 5 MSi to compare the change of phonon dispersion of NPMs in Figure 4 (a). Here, the c-Si$^{0.5\ MSi}$ pillar and c-Si$^{5\ MSi}$ pillar represent c-Si$^{M}$ pillar with M=0.5 MSi and 5 MSi, respectively, it is in the same way for a-Si$^{0.5\ MSi}$ pillar and a-Si$^{5\ MSi}$ pillar. The corresponding SED spectrum only including the atoms in Si membrane is shown in Figure 4 (b). The SED spectrum including all atoms in the NPMs is shown in Figure S4 in the Supplementary Material[42]. For M=0.5 MSi, there are fewer vibrational modes in NPMs with crystalline and amorphous pillars compared with NPMs with M=MSi (Figure 2 (a), Type 3 and Type 4). In addition, the frequency of the first hybridization moves from 0.15 THz in Figure 2 (a) for Type 3 and Type 4 to ~0.2 THz in Figure 4 (a) for M=0.5 MSi. These phenomena imply that the local resonant hybridization should be weakened, which in turn results in a larger $\kappa$ compared with NPM with M=MSi. For M=5 MSi, the frequency of the first hybridization moves to ~0.07 THz. However, the phonon dispersions from 0.7 to 1.4 THz are not severely changed as shown in Figure 4 (b), although the vibrational modes of the pillar are very dense.

### 3.3 The pillar height effect on the thermal conductivity of NPMs

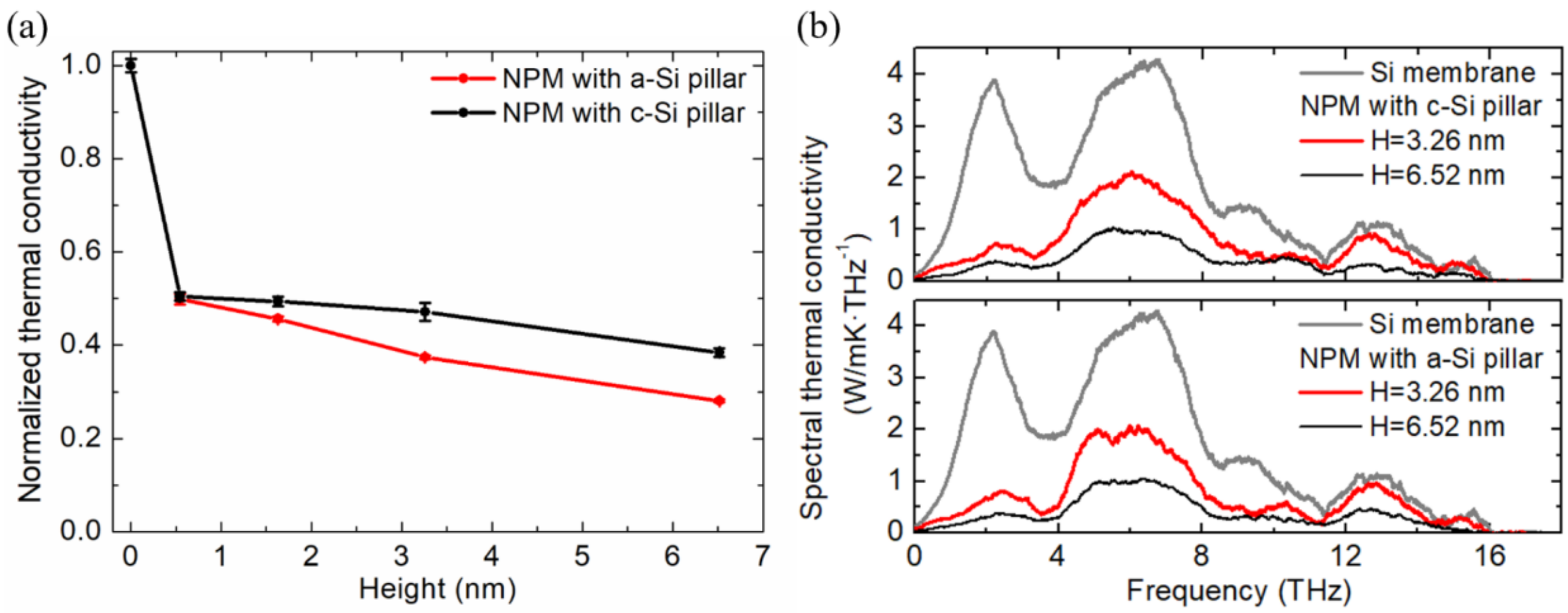

Figure 5 (a) Normalized thermal conductivity of NPM with c-Si pillar and a-Si pillar versus the height of pillars. (b) Spectral thermal conductivity of NPM with c-Si pillar and a-Si pillar for pillar heights of 3.26 nm and 6.52 nm.

Besides the different materials of pillars, the height (H) of pillars is another factor in tuning the $\kappa$ of NPMs.[9] Here, the pillars with height of H =0.54, 1.63, 3.26 and 6.52 nm are examined. The $\kappa$ of NPMs with c-Si pillar and a-Si pillar versus height are shown in Figure 5 (a). Further, the spectral $\kappa$ of NPMs are calculated in Figure 5 (b). For short pillars (H=0.54 nm), the $\kappa$ of NPMs with c-Si pillar is close to that of NPM with a-Si pillar, while the $\kappa$ is largely reduced by 61% for c-Si pillar and 72% for a-Si pillar as the height increases to 6.52 nm, which is consistent with previous reports[13,39]. Moreover, the a-Si pillars cause a larger reduction of $\kappa$ than the corresponding c-Si pillar for each pillar height case. Further analyses show that the spectral $\kappa$ of NPMs is reduced in the whole frequency range as the height of pillar increases in Figure 5 (b).

### 3.4 The thermal conductivity of NPMs with short pillars

Finally, the effect of the atomic mass on the $\kappa$ of NPMs with short pillars is investigated. The height of short pillar is 0.54 nm. The $\kappa$ of NPMs with short crystalline pillar and short amorphous pillar versus time is shown in Figure 6(a). Figure 6 (b) shows the $\kappa$ of NPM with short c-Si$^M$ pillar and short a-Si$^M$ pillar versus the value of M. It is found that the $\kappa$ is changed only ~4.3% for NPM with short c-Si$^M$ pillar as M increases from MSi to 10 MSi, while the $\kappa$ is changed 17% for NPM with c-Si$^M$ pillar in Figure 3(a). Therefore, the $\kappa$ of NPM with short c-Si$^M$ pillar is less sensitive to the change of the atomic mass, which means that the resonant hybridization mechanism is much weaker in NPMs with short crystalline pillar. Additionally, the $\kappa$ of NPM with short a-Si$^M$ pillar is almost unchanged as the value of M increases, which is the same as that in Figure 3 (a) for NPM with a-Si$^M$ pillar. These results indicate that resonant hybridization mechanism is weak on reducing the $\kappa$ of NPM with short c-Si pillar, and scatterings dominate on the reduction of $\kappa$ of both NPM with short c-Si pillar and NPM with short a-Si pillar.

Further, the phonon dispersions of NPMs with short pillars are calculated by lattice dynamics, and the SED spectrum of NPMs with short pillars only considering atoms in the Si membrane are also investigated, which are shown in Figure 6 (c) and (d). Compared with Figure 4 and Figure 2, there are smaller number of resonant hybridizations. This is because the short pillars have smaller number of atoms, thus smaller number of vibrational modes can be introduced. Therefore, hybridization mechanism is weaker in NPMs with short pillars compared with NPMs with higher pillars, which is consistent with the analyses of thermal conductivity in Figure 6 (a) and (b).

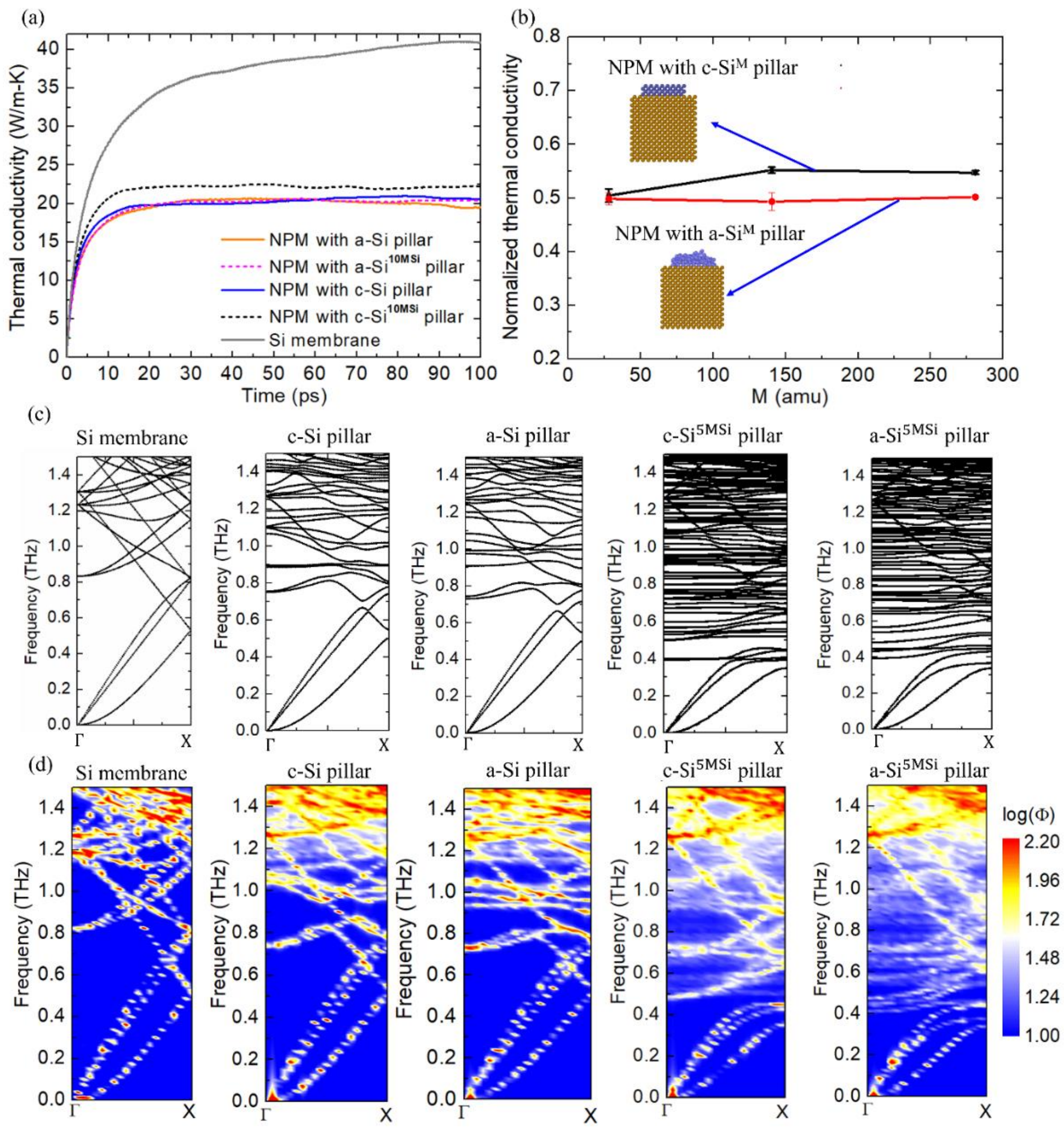

Figure 6 (a) Thermal conductivity of NPMs with short crystalline pillar and NPMs with short amorphous pillar versus time. The height of the pillars is H=0.54 nm. (b) Normalized thermal conductivity of NPM with short c-Si$^M$ pillar and short a-Si$^M$ pillar versus the value of M. The thermal conductivity of NPM is normalized by κ of the Si membrane. (c) Phonon dispersions of Si membrane and NPMs calculated by lattice dynamics. (d) SED spectrum of Si membrane and NPMs only considering atoms in the Si membrane. In (c) and (c), the M is set as 5 MSi for NPM with c-Si$^M$ pillar and NPM with a-Si$^M$ pillar.

Further, the difference between Si membrane with pillars and Si membrane with layers are investigated, which are shown in Figure S7 to S9 in the Supplementary Material[42]. The spatial distribution for pillars and layers is different. In NPMs, these pillars are discretely distributed, and the heat flux cannot directly transport between the adjacent pillars. However, the layers are continuously distributed on the surface of Si membrane, and the heat flux can directly transport in the layers. The difference of the spatial distribution of pillars and layers may cause the different effect on altering the thermal conductivity of Si membrane.

## 4. Conclusions

In this work, the thermal conductivity of NPM with crystalline Si pillar, crystalline Ge pillar, and amorphous Si pillar are systematically investigated by MD simulations. The phonon dispersion and spectral energy density show that the phonon dispersions are flattened due to local resonant hybridization induced by both the crystalline pillar and the amorphous pillar. In addition, a-Si pillar can cause larger reduction of thermal conductivity compared with c-Si pillar. Moreover, the thermal conductivity of NPM with crystalline Si pillar is increased as the atomic mass of atoms in the pillar increases because of the weakened resonant hybridization. However, the thermal conductivity of NPM with amorphous Si pillar is almost unchanged as the atomic mass of atoms in the pillar increases. The analyses of the reduction of thermal conductivity show that both resonant hybridization and scattering mechanism are important in NPM with crystalline pillar, while scattering mechanism dominates in NPM with amorphous pillar and NPM with short crystalline pillar. Further, the thermal conductivity of NPM is decreased as the pillar height increases for both crystalline pillar and amorphous pillar. The results of this work show that the thermal conductivity of NPMs can be tuned through the choice of pillar materials and heights for various applications requiring tailored thermal properties.

## 5. Acknowledgements

This work was sponsored by Beijing Institute of Technology Research Fund Program for Young Scholars (L.Y.).